\documentclass[12pt,preprint]{aastex}

\newcommand{\nc}{\newcommand}
\nc{\rlight}{r_L}
\nc{\be}{\begin{equation}}
\nc{\ee}{\end{equation}}
\nc{\bea}{\begin{eqnarray}}
\nc{\eea}{\end{eqnarray}}
\nc{\bfig}{\begin{figure}}
\nc{\efig}{\end{figure}}
\nc{\betabulk}{\beta_b}
\nc{\betadrift}{\beta_d}
\nc{\eperp}{e_\perp}
\nc{\eperppm}{e_\perp^\pm}
\nc{\epm}{e^\pm}
\nc{\Bbf}{{\bf B}}
\nc{\Ebf}{{\bf E}}
\nc{\bfu}{{\bf u}}
\nc{\Edimless}{E}
\nc{\Ex}{E_x}
\nc{\Ey}{E_y}
\nc{\Ez}{E_z}
\nc{\gam}{\gamma}
\nc{\gammaad}{\Gamma}
\nc{\incg}{\includegraphics}
\nc{\jump}{n_2/n_1}
\nc{\lft}{\left}
\nc{\rgt}{\right}
\nc{\rcrit}{r_c}
\nc{\lab}{\label}
\nc{\ldecay}{L_d}
\nc{\ltherm}{\Delta r_t}
\nc{\Lspin}{L}
\nc{\Lthirtyeight}{L_{38}}
\nc{\nddagger}{n_0}
\nc{\michelmu}{\mu}
\nc{\Ndot}{\dot N}
\nc{\Ndotthreeeight}{\dot N_{38}}
\nc{\Ndotforty}{\dot N_{40}}
\nc{\npm}{n^\pm}
\nc{\pperp}{p_\perp}
\nc{\pperppm}{p_\perp^\pm}
\nc{\rshock}{r_s}
\nc{\rshockfactor}{r_{s,15}}
\nc{\gammabulk}{{\gamma_0}}
\nc{\gammadrift}{{\gamma_d}}
\nc{\gammanull}{{\gamma_0}}
\nc{\gammapm}{\gamma^\pm}
\nc{\f}{\frac}
\nc{\omegacarrier}{\omega_0}
\nc{\omegaant}{\omegacarrier}
\nc{\omegaandrew}{\omega_\dagger}
\nc{\omeganull}{\omega_p}
\nc{\omegaopen}{\Omega_w}
\nc{\omegap}{\omega_p}
\nc{\omegaplasma}{{\omega_p}}
\nc{\omegaplasmaeff}{{\omega_p}_\dagger}
\nc{\omegaplasmadagger}{{\omega_{p\dagger}}}
\nc{\omegapd}{\omegaplasmadagger}
\nc{\nowtime}{\omegapd(t-\tref)}
\nc{\omegaspin}{\Omega}
\nc{\p}{\partial}
\nc{\pparav}{p_\|}
\nc{\pperpav}{p_\bot}
\nc{\uperp}{u_\bot}
\nc{\uperpthing}{\langle\uperp(\xdimless)\rangle/c}
\nc{\quiverlength}{\Delta l}
\nc{\sigmaant}{\Sigma_0}
\nc{\titlefreelowamp}{{\bf Low-Density Beam + High-$\sigma$ TEM Wave.\,}}
\nc{\titlefree}{{\bf High-Density Beam + High-$\sigma$ TEM Wave.\,}}
\nc{\titlesigmazero}{{\bf Ultrarelativistic Unmagnetized Shock.\,}}
\nc{\titlesigmazerofour}{{\bf Shocked High-Density Beam +
Moderate-$\sigma$ TEM Wave.\,}}
\nc{\titlesigmasix}{{\bf Shocked High-Density Beam + High-$\sigma$ TEM
Wave.\,}}
\nc{\Tpar}{T_\|}
\nc{\Tperp}{T_\bot}
\nc{\tref}{t_{\mbox{ref}}}
\nc{\tswitchon}{t_{\mbox r}}
\nc{\ttransit}{t_{\mbox c}}
\nc{\ux}{{u_x}}
\nc{\uy}{{u_y}}
\nc{\uz}{{u_z}}
\nc{\vshock}{v_s}
\nc{\xdimless}{\hat x}
\begin{document}

%%%%%%%%%%%%%%%%%%%%%%%%%%%%%%%%%%%%%%%%%%%%%%%%%%%%%%
%%%%%%%%%%%%%%%%%%%%%%%%%%%%%%%%%%%%%%%%%%%%%%%%%%%%%%

\title{Particle-In-Cell Simulations  
of a Nonlinear Transverse Electromagnetic Wave in a
Pulsar Wind Termination Shock}

\author{O. Skj\ae raasen}
\affil{Institute of Theoretical Astrophysics,
 University of Oslo, PO Box 1029 Blindern, N-0315 Oslo, Norway.}
\email{olafsk@astro.uio.no}
\author{A. Melatos}{
\affil{School of Physics, University of Melbourne,
Parkville, VIC 3010, Australia.}
\email{a.melatos@physics.unimelb.edu.au}
\author{A. Spitkovsky}
\affil{KIPAC, Stanford University, PO Box 20450, MS 29, Stanford, CA 94309.}
\affil{Chandra Fellow.}
\email{anatoly@slac.stanford.edu}

\begin{abstract}
A 2.5-dimensional particle-in-cell code is used to investigate
the propagation of a large-amplitude, superluminal, nearly transverse
electromagnetic (TEM) wave in a relativistically streaming electron-positron
plasma with and without a shock. 
In the freestreaming, unshocked case, the analytic TEM dispersion
relation is verified, and the streaming is shown to stabilize the wave against
parametric instabilities. In the confined, shocked case, the wave  
induces strong, coherent particle oscillations,
heats the plasma, and modifies the shock density profile via ponderomotive
effects. The wave decays over $\gtrsim 10^2$ skin depths; the decay length 
scale depends primarily on the ratio between the wave frequency and the
effective plasma frequency, and on the wave amplitude. The results are applied to  
the termination shock of the Crab pulsar wind, where the decay length-scale
$(\gtrsim 0.05\arcsec$) 
might be comparable to the thickness of filamentary, variable
substructure observed in the optical and X-ray wisps and knots.
\end{abstract}


\section{INTRODUCTION}
A rotation-powered pulsar emits most of its
spin-down luminosity as a wind of relativistic particles,
mainly electrons ($e^-$) and positrons ($e^+$) 
produced in pair cascades in charge-starved
regions of the magnetosphere \citep{arons02}. 
Observations of pulsar winds
confined by supernova remnants \citep{hesteretal02}
or the interstellar medium
\citep{chatterjee_cordes04,gaensler_etal04} 
reveal that the wind terminates in a synchrotron-emitting shock with a
cylindrically symmetric (`crossbow') morphology, implying that the outflow is 
collimated geometrically along the rotation axis, while the 
energy flux is concentrated 
in the equatorial plane \citep{komissarov_lyubarsky04}. 
The wind is launched from the light cylinder with most of its luminosity
transported as Poynting flux \citep{coroniti90}, and
its enthalpy flux is negligible beyond the light cylinder 
$\rlight$ due to adiabatic cooling.
On the other hand, models of the shock kinematics, the postshock flow,
the synchrotron spectrum \citep{kennelcoroniti84b}, and
the variability of shock substructure 
(e.g. wisps and knots in the Crab) \citep{sa04},
strongly suggest that the ratio $\sigma$ of Poynting flux to 
kinetic-energy flux is small at the shock, with $\sigma\sim 10^{-3}$
in several objects \citep{kennelcoroniti84a,gaensleretal02}.

In recent years, theoretical attention has focused on
the radial structure of the wind,
in order to explain how $\sigma$ decreases from the 
magnetosphere $(\sigma \gg 1)$ to the shock $(\sigma \ll 1)$:
the $\sigma$-paradox. 
The dissipation of Poynting flux seems to be related 
to the decay of the large-amplitude wave launched into the wind 
at the pulsar spin frequency.
One possible cause is that ideal magnetohydrodynamics (MHD)
breaks down beyond a critical radius $\rcrit$, in the sense
that there are insufficient charge carriers to screen out the rest-frame
electric field, and the electron inertia term in the relativistic Ohm's law
becomes dominant 
\citep{melatosmelrose96,melatos98,gedalin_etal01,kuijpers01,melatos02}.
In this respect, particular attention has been paid to the
entropy wave carrying the alternating (`striped') magnetic field  
in the wind \citep{bogovalov99,coroniti90} and its dissipation 
by magnetic reconnection at the corrugated current sheet separating 
the magnetic stripes \citep{lk01}, 
possibly emitting observable high-energy pulses 
\citep{arons79,kirketal02,kirk04,s04,petrikirk05}.
The dissipation occurs upstream from the termination shock, if
the reconnection rate and particle flux are
large enough \citep{ks03}, or else in the shock itself \citep{lyubarsky03}.

In this Letter, we assess an alternative scenario in which, beyond
$\rcrit$, the plasma upstream from the termination shock carries a
superluminal {\em transverse electromagnetic (TEM)} wave, which
modulates the wind and the shock \citep{melatos02}. The wave 
is taken to be monochromatic, with a frequency
given by the pulsar spin rate; an assumption generally used also in MHD wind models.
Coherent particle acceleration in the TEM wave
typically leads to $\sigma \ll 1$; a self-consistent, WKB wave model predicts
$\sigma \simeq 10^{-3}$
at the termination shock given the measured particle and energy fluxes of
the Crab pulsar \citep{melatosmelrose96,melatos98}.
We use the 2.5-dimensional, relativistic,
electromagnetic particle-in-cell (PIC) code XOOPIC \citep{vetal95}
to investigate the free propagation of the wave upstream from the shock
(\S 2), plasma heating in the shock (\S 3),
and the fate of the wave inside and downstream from the shock (\S 3). 
We apply our results to the Crab pulsar wind in \S 4.

%%%%%%%%%%%%%%%%%%%%%%%%%%%%%%%%%%%%%%%%%%%%%%%%%%%%%%%%%%%%%

\section{FREELY PROPAGATING TEM WAVES}
\lab{sec:noshock}
A nonlinear TEM wave can propagate even if its frequency $\omega$
is less than the effective plasma frequency 
$\sqrt{2\npm e^2/\epsilon_0m\gammapm}$,
where $\npm$ is the $\epm$ number density in the laboratory frame, 
$m$ is the electron mass, and $\gammapm$ is the
Lorentz factor, because the wave 
forces charges to oscillate relativistically, increasing their
effective mass. In a cold plasma, the plane wave dispersion relation is 
\citep{akhiezerpolovin1956,kawdawson1970,kennelpellat76,melatosmelrose96}
\bea
\eta^2 &=& 1 - \f{2\omeganull^2}{\omega^2\lft({1 + \Edimless^2}\rgt)^{1/2}},
\lab{eq:coldplasmanonlindisp}
\lab{eq:TEMnonlin_disp}
\lab{eq:one}
\eea
where $\eta = ck/\omega$ is the index of refraction,
$\omeganull = \sqrt{\npm e^2/\epsilon_0m\gammadrift}$,
$\gammadrift$ is the bulk Lorentz factor,
and $\Edimless$ is a dimensionless amplitude defined by
$\Edimless = {eE_0/ mc\omega}$, where $E_0$ is the physical electric field
amplitude. For $\Edimless >1$, the charges are
accelerated from nonrelativistic to relativistic speeds
within a single wave period. 
%The quantity $\omegaandrew$ is defined by
%\bea
%\omegaandrew^2 = \f{2\nddagger e^2}{m_e\epsilon_0}, &\,&
%n^\pm = \f{\nddagger}{1 - \beta_b\eta}, \lab{eq:labdensitypm}
%\eea
%where $c\beta_b$ is the plasma bulk speed. 
%For given values of $E_0$, $\gammabulk$, $\npm$ and
%$\omega$, one can solve (\ref{eq:TEMnonlin_disp}) and (\ref{eq:labdensitypm})
%for $k,\nddagger$ and $\omegaandrew$.
Note that a nonlinear electromagnetic wave is strictly transverse only if it is
circularly polarised. In this case, the streaming speed $c\betadrift$
is a constant of the motion and can be chosen arbitrarily.
As a result, $\sigma \propto E^2\gammadrift^{-1}$ 
varies independently with $\Edimless$ and $\gammadrift$, 
whereas in a linearly polarised
nonlinear wave, $\gammadrift$ (and hence $\sigma$) 
is determined uniquely by $\Edimless$. In the small-amplitude limit,
$\npm, \gammadrift$ and $\Edimless$ decouple.

In order to verify (\ref{eq:coldplasmanonlindisp}) with XOOPIC, 
we launch a cold $\epm$ beam from the left-hand edge 
($x = 0$) of the simulation box. 
Two orthogonal, phased dipole antennas are placed at 
$x=0$, continuously emitting a
circularly polarised TEM wave with frequency $\omega=\omegaant$ 
which ramps up to a constant 
amplitude $\Edimless$ over a rise time $\tswitchon \lesssim 0.2 \ttransit$,
where $\ttransit$ is the time for the beam to cross the box. 
The right-hand edge of the box is transparent both
to particles and waves (thanks to a wave-absorbing algorithm).
Once initial transients disappear, we Fourier transform the 
transverse electric field component $E_z(x,t)$ and measure
$\eta$ for the Fourier component of largest amplitude. 
For linear waves with $10 ^{-4} < \Edimless < 10^{-1}$, 
we verify the linearized version of (\ref{eq:coldplasmanonlindisp}) 
to approximately 0.5 per cent in vacuo, 
1 per cent in a uniform, cold stationary plasma, and 2 per cent for a
relativistically drifting, uniform, cold plasma. In the
evanescent regime $\omega < \omeganull\sqrt 2$, the wave penetrates a
few plasma skin depths $c/\omeganull$. 

When a TEM wave with $\Edimless \gg 1$ is launched into a nonstreaming
plasma, its ponderomotive force 
excavates a density cavity around the antenna. If the energy 
density of the wave field, $\epsilon_0E_0^2$, 
exceeds $\npm mc^2$, the cavity expands at relativistic speeds.
To verify (\ref{eq:coldplasmanonlindisp}) 
for a nonstreaming plasma, one would need to set up the 
wave field and particle distribution self-consistently 
at all $x$ before starting the
simulation (which we defer to a future paper),
and even then, one would expect the wave to be disrupted by parametric 
instabilities \citep{max_perkins72,sweeney_stewart75,ashour-abdallaetal81,leboeufetal82}.
However, if a nonlinear TEM wave
is launched into a streaming plasma ($\gammadrift \gg 1$),
it propagates freely --- without excavating the plasma and without
being disrupted by parametric instabilites ---
in the regime $\omega > \omeganull\sqrt 2/(1+E^2)^{1/4}$.
In the parameter ranges $1 < \Edimless < 10^3$ 
and  $1 < \gammadrift < 3.9\times 10^3$, the value of $\eta$ we compute
agrees with (\ref{eq:coldplasmanonlindisp}) to 5 per cent.
We stress, however, that more work is needed to rigorously 
verify that the simulated fields and particle momentum
distributions are compatible with the assumption of 
an infinite, plane TEM wave given by ({\ref{eq:coldplasmanonlindisp}).

\section{SHOCKED TEM WAVE}
In certain respects, the above PIC experiment with freely propagating
TEM waves overlaps with previous works
\citep{ashour-abdallaetal81,leboeufetal82}. The main focus of the current work
is a different scenario: A nonlinear TEM wave which encounters
an ultrarelativistic shock formed by a confining medium, as in the case of a
pulsar wind termination shock. In order to simulate such a shock, we
repeat the experiment described in $\S$2 but add a particle-reflecting,
wave-transmitting magnetic wall in the region 
$95 \lesssim \xdimless \equiv x\omegapd/c \lesssim 100$,
where $\omegapd = \omeganull(x=0)$. 
We trace O($10^6-10^7$) particles for O($10^3-10^4$) timesteps on a grid where
each cell has a size $\Delta x = \Delta y \lesssim 0.1c\pi/\omegapd$, and the 
timestep is given by $\Delta t = 0.4 \Delta x/c$. The grid has 128 cells and
periodic open boundaries in the $y$-direction, and from 256 to 1600 cells in the
$x$-direction. Below, the antenna (TEM wave) frequency is denoted by $\omega_0$.
We observe that the after reflected off the magnetic wall, 
the counterstreaming particles trigger a Weibel instability, which causes
the beam to filament in the $y$ direction. The initial stage
of the instability is marked by exponentially growing
oscillations (along the $y$ axis) of the magnetic field component $B_z$, 
with characteristic wavelength $c/\omegap$.
A shock forms after a time $t$ given by 
$\omegapd (t-\tref) \simeq 40 - 70$, where $\tref$ is the
instant when the front of the beam reflects off the magnetic
wall. As there is no dc magnetic
field in our simulations, the cyclotron instability seen 
by others \citep{gallantetal92,hoshinoetal92,sa04} is absent.

%%%done20050220

In the discussion below, we average all quantities over
$y$, smoothing out the filaments. We fix the bulk Lorentz 
factor at $\xdimless = 0$ to be $\gammanull = 3870$.
The value of $\sigma$ at the antenna, $\sigma(\xdimless=0)$, 
is denoted by $\sigmaant$, and we adopt parameters such that
$\Edimless(\xdimless = 0) = 5.4\times 10^3(\omegapd/\omegaant)\sqrt\sigmaant$. 
Downstream from the antenna, 
the field couples to the particles and quickly reduces $E$ until, 
at $\xdimless \approx 25$, 
a `fully developed' TEM wave 
forms, in which $E(\xdimless) \simeq \uperp(\xdimless)/c$, where 
$\uperp(\xdimless)$ is the local ensemble 
average of $(\uy^2+\uz^2)^{1/2}$ ($u_y$ and $u_z$  
are components of the 4-velocity).
We refer to the regions $10 \lesssim \xdimless \lesssim 50$, 
$50 \lesssim \xdimless \lesssim 80$,
and $\xdimless \gtrsim 80$ as the shock precursor, 
the shock interior and the downstream medium, respectively.

\subsection{Plasma Heating}
%%% BELOW: PHASE SPACE + SPECTRUM
In the shock precursor, there is a strong, coherent,
position-dependent interaction 
between the wave and the particles. The lower panels of Fig. \ref{fig:one} 
show the $e^+$ number density in $(\xdimless, u_z)$-space at  
two different positions. 
At low $\xdimless$, the particle motion is almost phase-coherent,
but electrostatic 
and electromagnetic fluctuations gradually increase the thermal spread
with $\xdimless$ until it exceeds the wave-induced quiver motion.
The conversion of Poynting flux and 
kinetic-energy flux into thermal motions is evident 
from the upper panels of Fig. \ref{fig:one}, which show the frequency
spectrum of the electric field component $E_y$. As $\xdimless$ increases,
the thermal background (broad sidebands) grows at the expense of the
antenna-driven TEM mode (narrow lines at $\omega=\pm\omegaant$).

%%% BELOW: LORENTZ FACTOR DISTN
The positron energy distribution $f(\gam)$ 
is shown at two positions in Fig. \ref{fig:gammadist}, one in
the shock precursor and the other in the shock interior.
Interestingly, for a wave with $\sigmaant \gtrsim 1$, 
we find $\langle \gamma \rangle \propto \gammanull(\sigmaant + 1)$ at both positions
(indeed, throughout the shock), although the form of $f(\gamma)$ changes. 
In the precursor ($8 \lesssim \xdimless \lesssim 16$), the distribution 
is narrow and drops quickly at high energies;
fitting the tail with a power law, $f(\gamma) \simeq \gamma^{-\alpha}$,
we obtain $15 \lesssim \alpha \lesssim 35$
for the range $2 \lesssim \sigmaant \lesssim 24$.
This does not necessarily imply nonthermal acceleration, 
but rather the heating of an initially cold distribution. 
In the shock interior ($50 \lesssim \xdimless \lesssim 58$), 
$f(\gamma)$ turns into 
an anisotropic Maxwellian (see $\S$\ref{sec:densityprofile}). 
The stronger the TEM wave, the hotter the downstream plasma, 
with $\langle \gamma \rangle$ 
determined by the {\it total} upstream energy flux.

The anisotropy of the particle distribution in the shock interior
depends on the polarization of the wave; for example, a linearly
polarized wave with $E_z \gg 1$ and $E_y = 0$ predominantly energizes $\uz$.
A detailed study of this issue is outside the scope of this paper,
which focuses on circularly polarized TEM waves.

%%%done20050220

%%%%%%%%%%%%%%%%%%%%%%%%%%%%%%%%%%%%%%%%%%%%%%%%%%%%%%%%%%%%%%%%%%%
%%% BELOW: TEM WAVE DECAY
\subsection{Fate of the Wave}
\lab{sec:ponderomotive}
As the wave propagates 
away from the antenna and accelerates the particles into oscillations,
$\sigma$ quickly drops below $\sigmaant$.
For $\xdimless \gtrsim 25$, $E$ changes little until it encounters the shock,
beyond which the wave continues to decay.
The decay rate scales with the skin depth, $c/\omegapd$. 
Figure \ref{fig:ldecay}a
shows the TEM wave amplitude as a function of position and $\sigmaant$.
For $\sigmaant = 0.4$, we find 
$E(\xdimless=28)/E(\xdimless=0)= 0.05$ and 
$E(\xdimless=40)/E(\xdimless=0)\lesssim 0.01$.
For $\sigmaant = 3.6\, (14.0)$, the corresponding ratios
are $0.20\, (0.25)$ and $0.07\, (0.14)$, respectively.
In the case $\sigmaant = 14.0$, $E$
remains roughly flat for $40 < \xdimless < 70$, 
before dropping away for $\xdimless \gtrsim
70$. The drop is accompanied by a temperature decline, possibly
because the plasma near the magnetic wall is not yet thermalized.
For $\sigmaant \simeq 6$, a similar drops occurs at $\xdimless \gtrsim 57$.

Taken together with similar simulations spanning
the parameter range $1 \lesssim \omegaant/\omegapd \lesssim 10$, 
Fig. 3a leads to an important result: the  
length scale $\ldecay$ over which $E$ drops by two orders of magnitude,
scales proportionally to the skin depth as
\bea
\ldecay &\simeq& K c/\omegapd,
\eea 
where $10^2\lesssim K\lesssim 10^3$ depends on $\omegaant/\omegapd$; i.e., near
(far from) from the cutoff given by (1) the decay is faster (slower). This is
consistent with previous results for an unshocked, relativistic EM wave
\citep{leboeufetal82}. The value of $K$ also increases slowly with the amplitude
(for $0.2 < \sigmaant < 20$ we obtain $ 50 \lesssim K \lesssim 150$ with
$\omegaant/\omegapd \simeq 2.0$). For the parameter regime 
$\omegaant/\omegapd \gg 10$, which we cannot achieve in our simulations (since it
would require a very large number of timesteps to both allow the shock to
form and resolve the wave oscillations), the value of $K$ might exceed $10^3$. 
Over distances much greater than $\ldecay$, the wave may enter the linear regime,
but we are unable to distinguish the wave from the noise at this level 
in our simulations. (Note that if $\ldecay$ is redefined as the length scale over which
$E$ drops by a factor of $e$, $K$ is reduced by a factor of 4.6).

As seen from Fig. \ref{fig:ldecay}b, 
as the  wave decays the tranverse 
momentum density $\eperp(\xdimless) = n(\xdimless)\uperp(\xdimless)/c$,
where $n(\xdimless) = n^+(\xdimless) + n^-(\xdimless)$,
increases with $\xdimless$. The largest-amplitude
case ($\sigmaant = 14.0$) is an exception, with $\p\eperp/\p\xdimless < 0$
near $\xdimless = 0$ and the magnetic wall;
this may be due to ponderomotive forces (\S \ref{sec:densityprofile}),
imperfect absorption of the wave by the simulation boundaries, or other
boundary effects.

The wave decay might be caused by a parametric instability which sets in
when the bulk speed of the flow is reduced in the shock interior. Further
simulations are needed to obtain enough statistics to 
rigorously settle the microphysical aspects of the
decay; this is deferred to a forthcoming work.

\subsection{Density Profile}
%%% BELOW: DENSITY PROFILE
\lab{sec:densityprofile}
Our code tracks all three vector components of the electromagnetic 
fields, but only supports electrostatic (Langmuir) fluctuations 
in two directions. Particles therefore diffuse slower
in $u_z$ than in $u_x$ and $u_y$.
In this respect, our simulations are 
essentially two-dimensional, with an effective adiabatic index 
$\gammaad \simeq 3/2$ (not 4/3, as in isotropic 3D).
This explains the form of the density profile $n(x)$ in Fig. \ref{fig:four},
plotted at $t = \tref + 85\omegapd^{-1}$ for three different wave amplitudes.
In the near-hydrodynamic case $\sigmaant \simeq 0$ (not shown), 
the density jump is $\jump \simeq 2.8$, where $n_1$ ($n_2$) denotes the 
upstream (downstream) density, consistent with 
the Rankine-Hugoniot prediction
$\jump = \gammaad/(\gammaad - 1) \simeq 3$ for $\gammaad = 1.5$.
Conversely, given the observed density jump for $\sigmaant\simeq 0$, 
we infer $\gammaad = 1.56$,
implying some slow diffusion in $u_z$.
Figure \ref{fig:four} also suggests that 
$n_2/n_1$ increases with $\sigmaant$.
This trend can be ascribed to the ponderomotive force executed by the decaying
TEM wave, and/or by the longitudinal electric field $\Ex \propto E$
induced by parametric decay to the Langmuir mode.
As the TEM wave decays, we have 
a ponderomotive force proportional to  $\p E^2/\p x$;
hence $n_2/n_1$ increases with $\sigmaant$.
Interestingly, we see little or 
no evidence that the shock speed changes with $\sigmaant$.

%%%%%%%%%%%%%%%%%%%%%%%%%%%%%%%%%%%%%%%%%%%%%%%%%%%%%%%%%%%

\section{CRAB PULSAR WIND}
For a steady-state pulsar wind, conservation
of mass and energy give
\bea
{\omegaant\over\omeganull} &=& 
4.6\times 10^{-8}
\lft(\f{r}{\rlight}\rgt)
\lft[{\gammadrift\lft(\omegaopen/4\pi \rgt)
\over \Ndotthreeeight}\rgt]^{1/2}, \lab{eq:four}\\
\Edimless &=& 3.1\times 10^7
\lft(\f{\rlight}{r}\rgt)\lft[
\f{\Ndotthreeeight\michelmu\sigma}{\eta\lft(1 + \sigma\rgt)
\lft(\omegaopen/4\pi\rgt)}
\rgt]^{1/2}. \lab{eq:five}
\eea
In (\ref{eq:four}) and (\ref{eq:five}), $\omegaant$ is given by 
the pulsar spin rate, the total particle injection rate is 
$\Ndot = 10^{38}s^{-1}\Ndotthreeeight$, the solid angle filled by the
wind is $\omegaopen$, and we define $\michelmu = L/\Ndot mc^2$, where $L$
is the pulsar spin-down luminosity. For canonical Crab parameters
($5\times 10^4 \lesssim \michelmu \lesssim 5\times 10^6$;
$\eta \simeq 1$,
$\omegaopen/4\pi \simeq 0.1 - 1$, $\sigma(\rshock)\simeq 3\times 10^{-3}$, 
and $\rshock/\rlight \simeq 2.7 \times 10^9$)
\citep{trumperbecker98,weisskopfetal00,ks03,sa04},
we take $\omegapd = \omeganull(\rshock)$ and 
find $1 \lesssim \Edimless \lesssim 10$,
$9\times 10^2 \lesssim \omegaant/\omegapd \lesssim 3 \times 10^5$,
and $9\times 10^4 \lesssim \ldecay/\rlight \lesssim 3\times 10^8$ 
at the termination shock (using $10^2\lesssim K \lesssim 10^3$). 

In a self-consistent model of the TEM wave, $\sigma$ can be expressed 
in terms of the other quantities in (1), (3), and (4) according to Eq. (97) of 
\citet{melatosmelrose96}, viz. 
\bea
\sigma &=& \f{\eta}{\betadrift}
\lft[\f{1}{\gammadrift^2\lft(1-\eta^2\rgt)} - 
\f{\omega^4\lft(1-\eta^2\rgt)}{4\gammadrift^2\omeganull^4}\rgt].
\eea
This allows us to solve self-consistently for $\eta, \sigma, \Edimless$, and
$\omegapd$ given just $\Ndot$ and
$\gammadrift$. Adopting the current best estimates of these quantities,
$\Ndotthreeeight=10^{0.3}$ and $\gammadrift=10^6$, 
as inferred from an ion-cyclotron model of the variability of the Crab wisps 
\citep{sa04}, together with $\omegaopen \simeq 4\pi$, we solve (1) and (3)--(5) to obtain 
$\sigma\simeq 9.0\times 10^{-3}$, $E\simeq 2.5$, 
$\ldecay/\rlight \simeq 10^7 - 10^8$, and
$1-\eta \simeq 4.8\times 10^{-11}$
(the value for $\eta$ reflects the fact that the 
displacement current dominates the conduction current for $r>\rcrit$).

Evidently, our simulations explore artificially small values of
$\omegaant/\omegapd$ (for best-guess values, $\omegaant/\omegapd \approx 9
\times 10^4$ at the Crab shock), and
artificially large values of $\sigma$ and $\Edimless$ (e.g.,
in Fig. 4, $\sigmaant = 3.6$ gives $\sigma \simeq 0.01$ 
and $\Edimless \simeq 4\times 10^2$ at $\xdimless \simeq 50$) due to 
numerical limitations. Nevertheless, we consistently find
$\ldecay \gtrsim 10^2c/\omegapd$ for 
$0.1 \lesssim \sigmaant \lesssim 100$, i.e.,
over three decades in $\sigmaant$. The fate of the
TEM wave thus seems clear: It decays beyond $\rshock$, 
just like the entropy wave \citep{lyubarsky03}, but the dissipation mechanism
is not reconnection. The decay scale exceeds $10^7\,\rlight$.

Observationally, $\ldecay$ might be resolved. 
It subtends $0.05\arcsec-0.5\arcsec$
if the Crab is at distance of 2 kpc, using $10^2<K<10^3$ and 
the best estimates for the wind parameters. This 
range includes the angular thickness of filamentary, variable substructure
in the equatorial wisps of the Crab pulsar wind, observed by the 
{\it Hubble Space Telescope} and the {\it Chandra X-ray Observatory} 
\citep{hesteretal02}, but at present the connection is speculative. Since
$\ldecay$ increases with $\omega/\omegapd$ in the parameter regime
near cutoff, and the Crab shock value 
of $\omega/\omegapd$ cannot be achieved in our simulations, 
we can only conclusively give a
{\it lower limit} ($100c/\omegapd$) for $\ldecay$; i.e., the possibility 
that $\ldecay > 10^3c/\omegapd$ ($>0.5\arcsec$) cannot be excluded.
A theoretical upper limit for $\ldecay$ can be estimated by noting that, ultimately,
adiabatic cooling of the diverging, subrelativistic flow beyond the Crab termination shock
enforces a cutoff, 
and parametric instabilities are bound to set in, yielding $\ldecay \lesssim O(\rshock)$.

The TEM wave generates a latitude-dependent, nonthermal particle distribution
inside the shock which is anisotropic near the equatorial plane,
where the wave is (approximately) linearly polarized, and isotropic 
at high latitudes, where the polarization is (approximately) 
circular. In principle, future X-ray and/or optical polarization measurements 
can test this prediction. 

The simulations reported here do not include a dc magnetic field,
as exists for an oblique rotator.
Such a field affects the propagation and stability of the TEM wave
\citep{asseoetal1980}.
For example, a subluminal, circularly polarized TEM wave with a dc
magnetic field in the $x$-direction (radially) is stable in the inner part
of the wind but unstable to three-wave parametric decay in the outer wind 
\citep{melatos98,melatos02}. 
These important issues will be addressed in a forthcoming paper.

\acknowledgments{O.S. thanks the Astrophysics Group, University of Melbourne,
the Norwegian Research Council, 
and Prof. Jean Heyvaerts, Observatoire de Strasbourg, for all support.
A.S. acknowledges support provided by NASA through Chandra Fellowship
grant PF2-30025 awarded by the Chandra X-Ray Center, which is operated
by the Smithsonian Astrophysical Observatory for NASA under contract
NAS8-39073. }

\bibliographystyle{apj}
\bibliography{references}

\begin{thebibliography}{37}
\expandafter\ifx\csname natexlab\endcsname\relax\def\natexlab#1{#1}\fi

\bibitem[{{Akhiezer} \& {Polovin}(1956)}]{akhiezerpolovin1956}
{Akhiezer}, A.~I., \& {Polovin}, R.~V. 1956, Sov. Phys. JETP, 3, 696

\bibitem[{{Arons}(1979)}]{arons79}
{Arons}, J. 1979, Space Science Reviews, 24, 437

\bibitem[{{Arons}(2002)}]{arons02}
{Arons}, J. 2002, in Astronomical Society of the Pacific Conference Series,
  Vol. 271, 71

\bibitem[{{Ashour-Abdalla} {et~al.}(1981){Ashour-Abdalla}, {Leboeuf}, {Tajima},
  {Dawson}, \& {Kennel}}]{ashour-abdallaetal81}
{Ashour-Abdalla}, M., {Leboeuf}, J.~N., {Tajima}, T., {Dawson}, J.~M., \&
  {Kennel}, C.~F. 1981, \pra, 23, 1906

\bibitem[{{Asseo} {et~al.}(1980){Asseo}, {Llobet}, \&
  {Schmidt}}]{asseoetal1980}
{Asseo}, E., {Llobet}, X., \& {Schmidt}, G. 1980, \pra, 22, 1293

\bibitem[{{Bogovalov}(1999)}]{bogovalov99}
{Bogovalov}, S.~V. 1999, \aap, 349, 1017

\bibitem[{{Chatterjee} \& {Cordes}(2004)}]{chatterjee_cordes04}
{Chatterjee}, S., \& {Cordes}, J.~M. 2004, \apjl, 600, L51

\bibitem[{{Coroniti}(1990)}]{coroniti90}
{Coroniti}, F.~V. 1990, \apj, 349, 538

\bibitem[{{Gaensler} {et~al.}(2002){Gaensler}, {Arons}, {Kaspi}, {Pivovaroff},
  {Kawai}, \& {Tamura}}]{gaensleretal02}
{Gaensler}, B.~M., {Arons}, J., {Kaspi}, V.~M., {Pivovaroff}, M.~J., {Kawai},
  N., \& {Tamura}, K. 2002, \apj, 569, 878

\bibitem[{{Gaensler} {et~al.}(2004){Gaensler}, {van der Swaluw}, {Camilo},
  {Kaspi}, {Baganoff}, {Yusef-Zadeh}, \& {Manchester}}]{gaensler_etal04}
{Gaensler}, B.~M., {van der Swaluw}, E., {Camilo}, F., {Kaspi}, V.~M.,
  {Baganoff}, F.~K., {Yusef-Zadeh}, F., \& {Manchester}, R.~N. 2004, \apj, 616,
  383

\bibitem[{{Gallant} {et~al.}(1992){Gallant}, {Hoshino}, {Langdon}, {Arons}, \&
  {Max}}]{gallantetal92}
{Gallant}, Y.~A., {Hoshino}, M., {Langdon}, A.~B., {Arons}, J., \& {Max}, C.~E.
  1992, \apj, 391, 73

\bibitem[{{Gedalin} {et~al.}(2001){Gedalin}, {Gruman}, \&
  {Melrose}}]{gedalin_etal01}
{Gedalin}, M., {Gruman}, E., \& {Melrose}, D.~B. 2001, \mnras, 325, 715

\bibitem[{{Hester} {et~al.}(2002){Hester}, {Mori}, {Burrows}, {Gallagher},
  {Graham}, {Halverson}, {Kader}, {Michel}, \& {Scowen}}]{hesteretal02}
{Hester}, J.~J., {Mori}, K., {Burrows}, D., {Gallagher}, J.~S., {Graham},
  J.~R., {Halverson}, M., {Kader}, A., {Michel}, F.~C., \& {Scowen}, P. 2002,
  \apjl, 577, L49

\bibitem[{{Hoshino} {et~al.}(1992){Hoshino}, {Arons}, {Gallant}, \&
  {Langdon}}]{hoshinoetal92}
{Hoshino}, M., {Arons}, J., {Gallant}, Y.~A., \& {Langdon}, A.~B. 1992, \apj,
  390, 454

\bibitem[{{Kaw} \& {Dawson}(1970)}]{kawdawson1970}
{Kaw}, P., \& {Dawson}, J. 1970, Phys. Fluids, 13, 472

\bibitem[{{Kennel} \& {Coroniti}(1984{\natexlab{a}})}]{kennelcoroniti84a}
{Kennel}, C.~F., \& {Coroniti}, F.~V. 1984{\natexlab{a}}, \apj, 283, 694

\bibitem[{{Kennel} \& {Coroniti}(1984{\natexlab{b}})}]{kennelcoroniti84b}
---. 1984{\natexlab{b}}, \apj, 283, 710

\bibitem[{{Kennel} \& {Pellat}(1976)}]{kennelpellat76}
{Kennel}, C.~F., \& {Pellat}, R. 1976, J. Plasma Phys., 15, 335

\bibitem[{{Kirk}(2004)}]{kirk04}
{Kirk}, J.~G. 2004, Physical Review Letters, 92, 181101

\bibitem[{{Kirk} \& {Skj{\ae}raasen}(2003)}]{ks03}
{Kirk}, J.~G., \& {Skj{\ae}raasen}, O. 2003, \apj, 591, 366

\bibitem[{{Kirk} {et~al.}(2002){Kirk}, {Skj{\ae}raasen}, \&
  {Gallant}}]{kirketal02}
{Kirk}, J.~G., {Skj{\ae}raasen}, O., \& {Gallant}, Y.~A. 2002, \aap, 388, L29

\bibitem[{{Komissarov} \& {Lyubarsky}(2004)}]{komissarov_lyubarsky04}
{Komissarov}, S.~S., \& {Lyubarsky}, Y.~E. 2004, \mnras, 349, 779

\bibitem[{{Kuijpers}(2001)}]{kuijpers01}
{Kuijpers}, J. 2001, Publications of the Astronomical Society of Australia, 18,
  407

\bibitem[{{Leboeuf} {et~al.}(1982){Leboeuf}, {Ashour-Abdalla}, {Tajima},
  {Kennel}, {Coroniti}, \& {Dawson}}]{leboeufetal82}
{Leboeuf}, J.~N., {Ashour-Abdalla}, M., {Tajima}, T., {Kennel}, C.~F.,
  {Coroniti}, F.~V., \& {Dawson}, J.~M. 1982, \pra, 25, 1023

\bibitem[{{Lyubarsky} \& {Kirk}(2001)}]{lk01}
{Lyubarsky}, Y., \& {Kirk}, J.~G. 2001, \apj, 547, 437

\bibitem[{{Lyubarsky}(2003)}]{lyubarsky03}
{Lyubarsky}, Y.~E. 2003, \mnras, 345, 153

\bibitem[{{Max} \& {Perkins}(1972)}]{max_perkins72}
{Max}, C., \& {Perkins}, F. 1972, Physical Review Letters, 29, 1731

\bibitem[{{Melatos}(1998)}]{melatos98}
{Melatos}, A. 1998, Memorie della Societa Astronomica Italiana, 69, 1009

\bibitem[{{Melatos}(2002)}]{melatos02}
{Melatos}, A. 2002, in Astronomical Society of the Pacific Conference Series,
  Vol. 271, 115

\bibitem[{{Melatos} \& {Melrose}(1996)}]{melatosmelrose96}
{Melatos}, A., \& {Melrose}, D.~B. 1996, \mnras, 279, 1168

\bibitem[{{P{\' e}tri} \& {Kirk}(2005)}]{petrikirk05}
{P{\' e}tri}, J., \& {Kirk}, J.~G. 2005, \apjl, 627, L37

\bibitem[{{Skj{\ae}raasen}(2004)}]{s04}
{Skj{\ae}raasen}, O. 2004, Advances in Space Research, 33, 586

\bibitem[{{Spitkovsky} \& {Arons}(2004)}]{sa04}
{Spitkovsky}, A., \& {Arons}, J. 2004, \apj, 603, 669

\bibitem[{{Sweeney} \& {Stewart}(1975)}]{sweeney_stewart75}
{Sweeney}, G.~S.~S., \& {Stewart}, P. 1975, \aap, 41, 431

\bibitem[{{Tr{\" u}mper} \& {Becker}(1998)}]{trumperbecker98}
{Tr{\" u}mper}, J., \& {Becker}, W. 1998, Advances in Space Research, 21, 203

\bibitem[{{Verboncoeur} {et~al.}(1995){Verboncoeur}, Langdon, \&
  {Gadd}}]{vetal95}
{Verboncoeur}, J.~P., Langdon, A.~B., \& {Gadd}, N.~T. 1995, Comp. Phys. Comm.,
  87, 199

\bibitem[{{Weisskopf} {et~al.}(2000){Weisskopf}, {Hester}, {Tennant}, {Elsner},
  {Schulz}, {Marshall}, {Karovska}, {Nichols}, {Swartz}, {Kolodziejczak}, \&
  {O'Dell}}]{weisskopfetal00}
{Weisskopf}, M.~C., {Hester}, J.~J., {Tennant}, A.~F., {Elsner}, R.~F.,
  {Schulz}, N.~S., {Marshall}, H.~L., {Karovska}, M., {Nichols}, J.~S.,
  {Swartz}, D.~A., {Kolodziejczak}, J.~J., \& {O'Dell}, S.~L. 2000, \apjl, 536,
  L81

\end{thebibliography}

\clearpage

\bfig
\incg[width=0.49\textwidth]{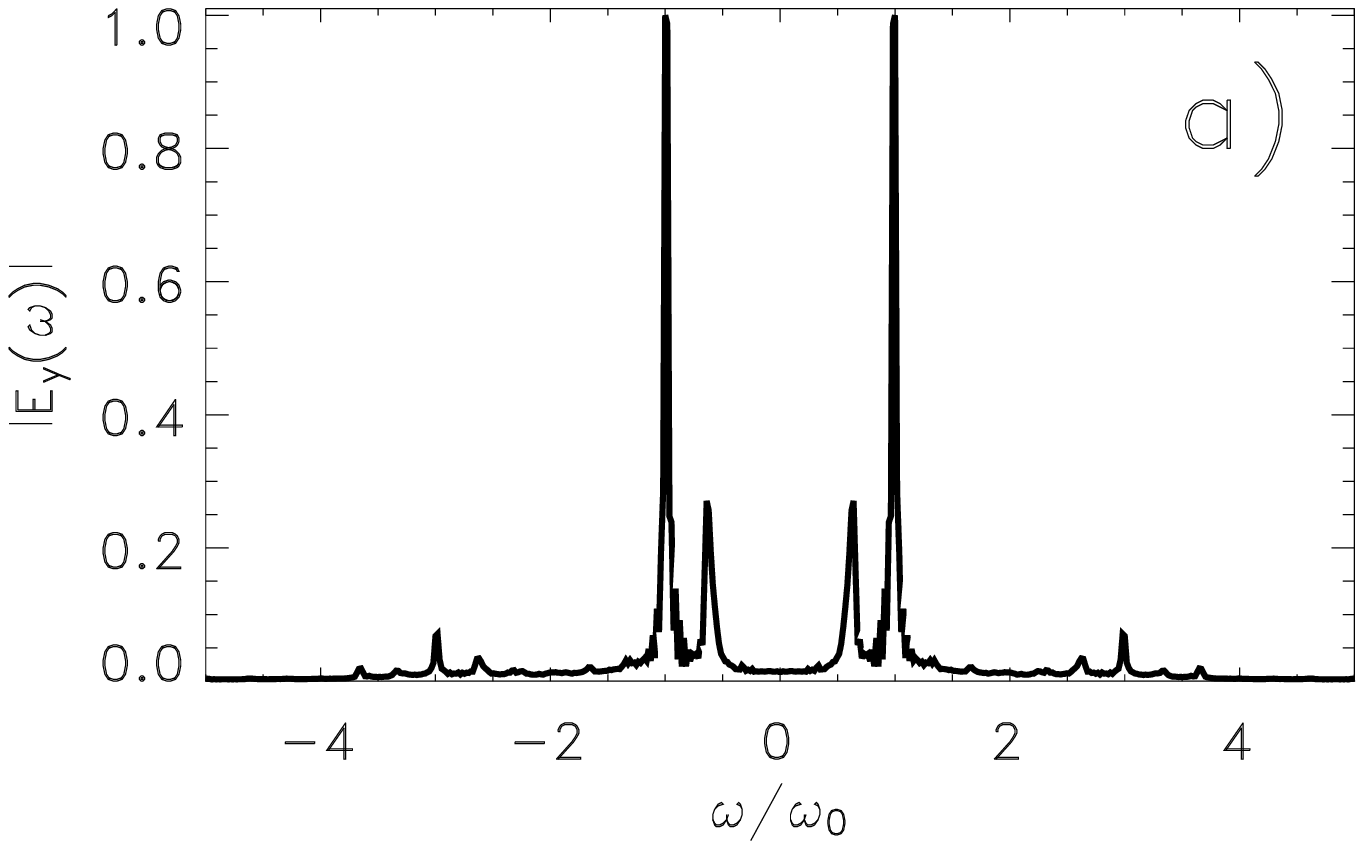}
\incg[width=0.49\textwidth]{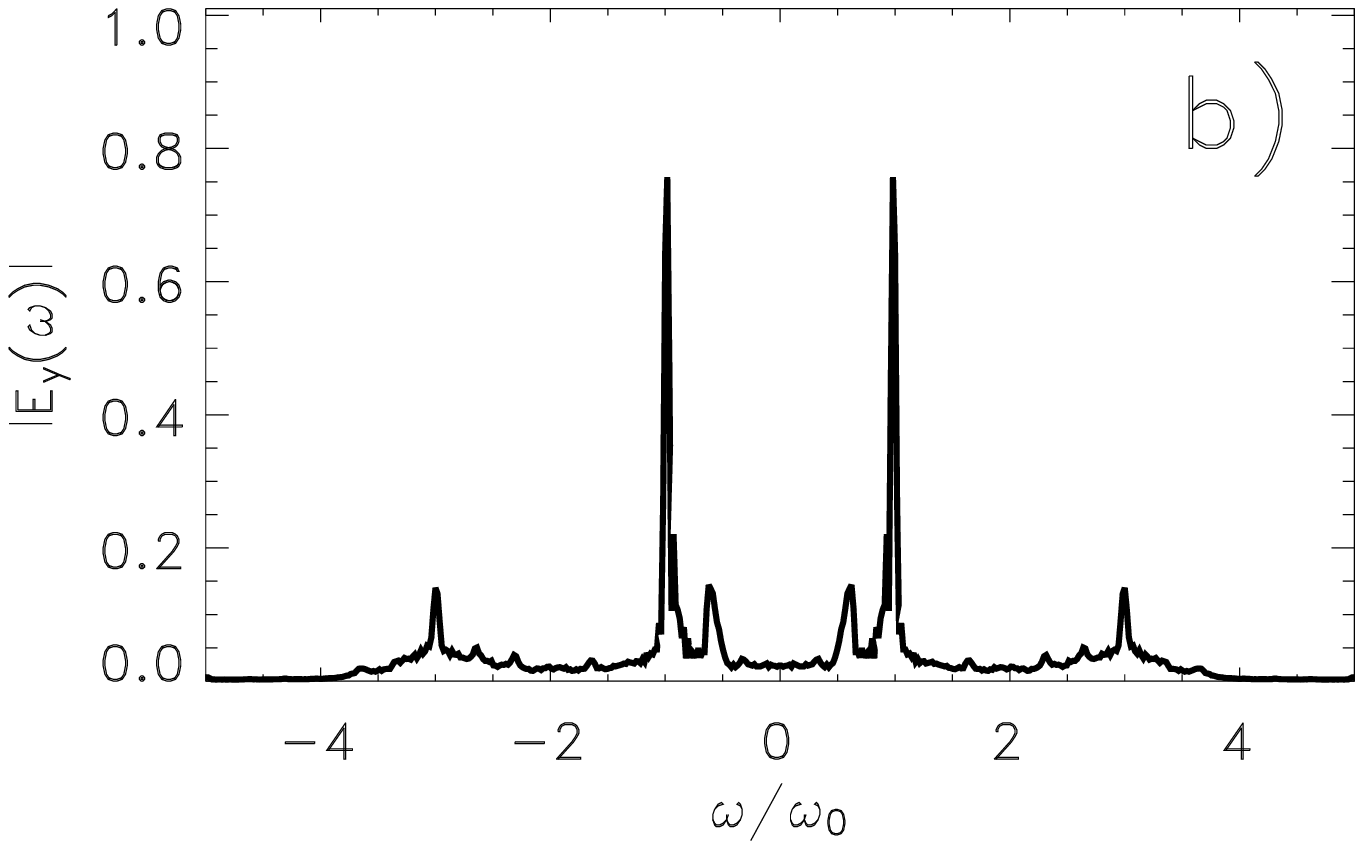}
\newline
\incg[width=0.49\textwidth]{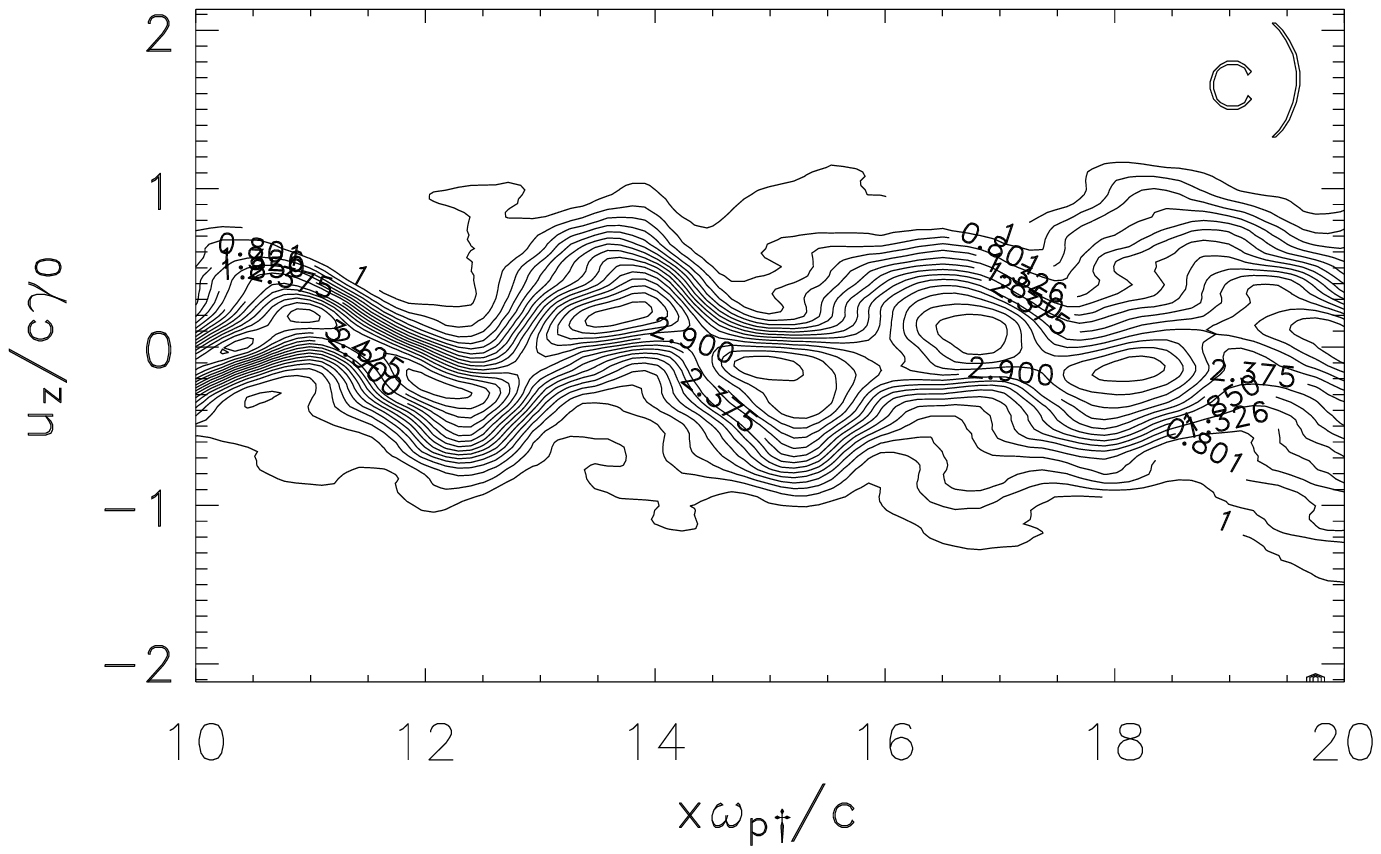}
\incg[width=0.49\textwidth]{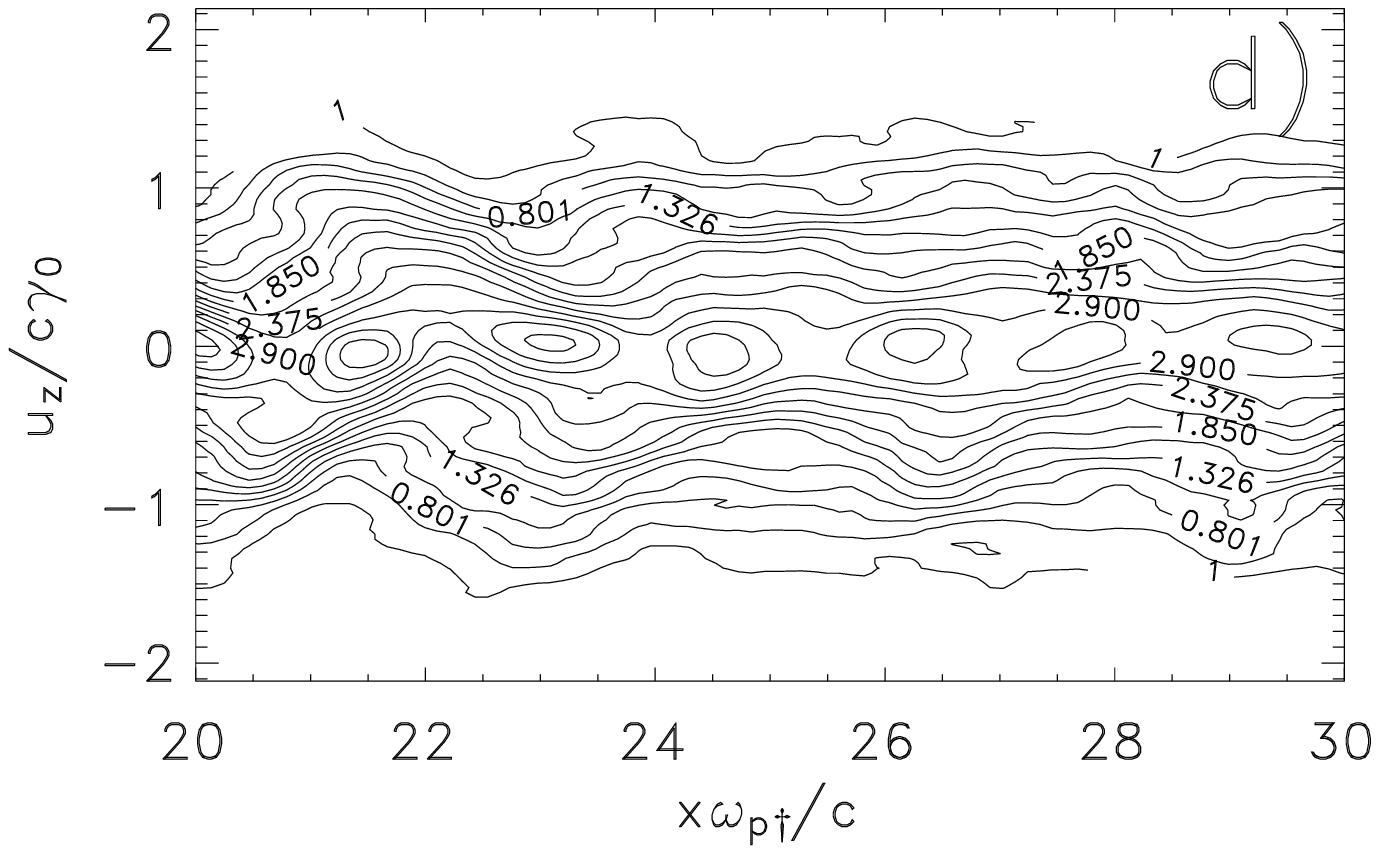}
\caption{The shock precursor. {\it Top row}: 
Frequency spectrum $|E_y(\omega)|$ at $(a)$ $\xdimless = 14$ and
$(b)$ $\xdimless = 29$, obtained during the time interval   
$56 \lesssim \omegapd (t-\tref) \lesssim 80$. The lines  
at $\omega = \pm\omegaant$ are the nonlinear TEM wave. 
As $\xdimless$ increases, the lines decay and broaden while the 
thermal background spectrum grows. The lines at 
$\omega\simeq \pm 0.6\omegaant$ are due to electrostatic plasma oscillations.
{\it Bottom row}: Positron density in $(x,\uz)$-space for 
$(c)$ $10 < \xdimless < 20$ and $(d)$ $20 < \xdimless < 30$. 
In (c), the particles 
are strongly phase-coherent with the wave 
(which has a wavelength of $\approx 3.3\xdimless$), 
although thermal broadening can be seen at $\xdimless > 11$. 
In (d), the thermal width is larger than the
quiver amplitude, but a periodic modulation is still evident
(in the absence of the wave, one gets a straight
beam centered on $\uz = 0$, gradually broadening with $\xdimless$).
The parameters used are $\sigmaant = 2.0$,
$\gammabulk = 3870$, $\omegaant/\omegapd = 1.97$,
$t=\tref + 85/\omegapd$, and
$E(\xdimless = 0) = 2.0\times 10^3\sqrt\sigmaant$.}
\lab{fig:uz}
\lab{fig:one}
\efig

\bfig
\incg[width=0.5\textwidth]{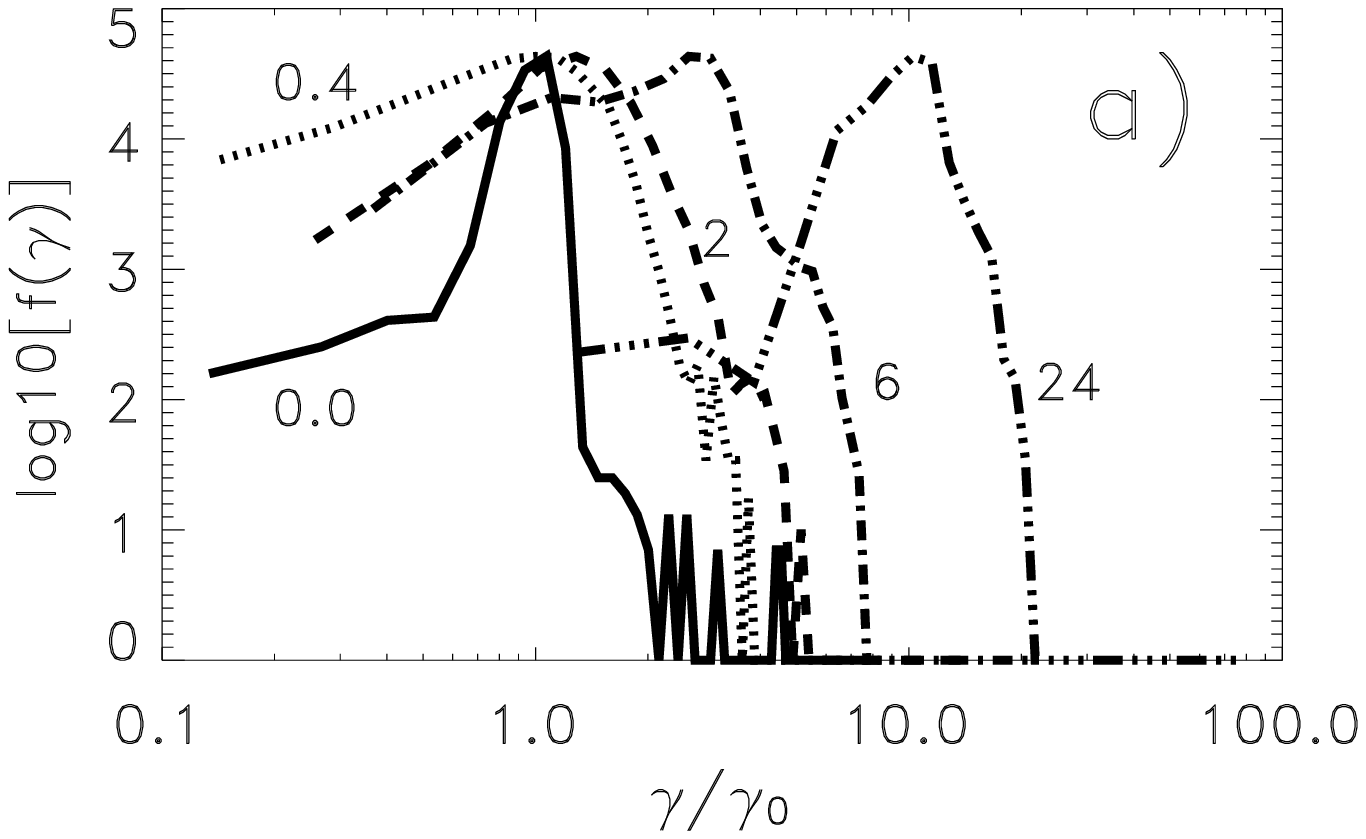}
\incg[width=0.5\textwidth]{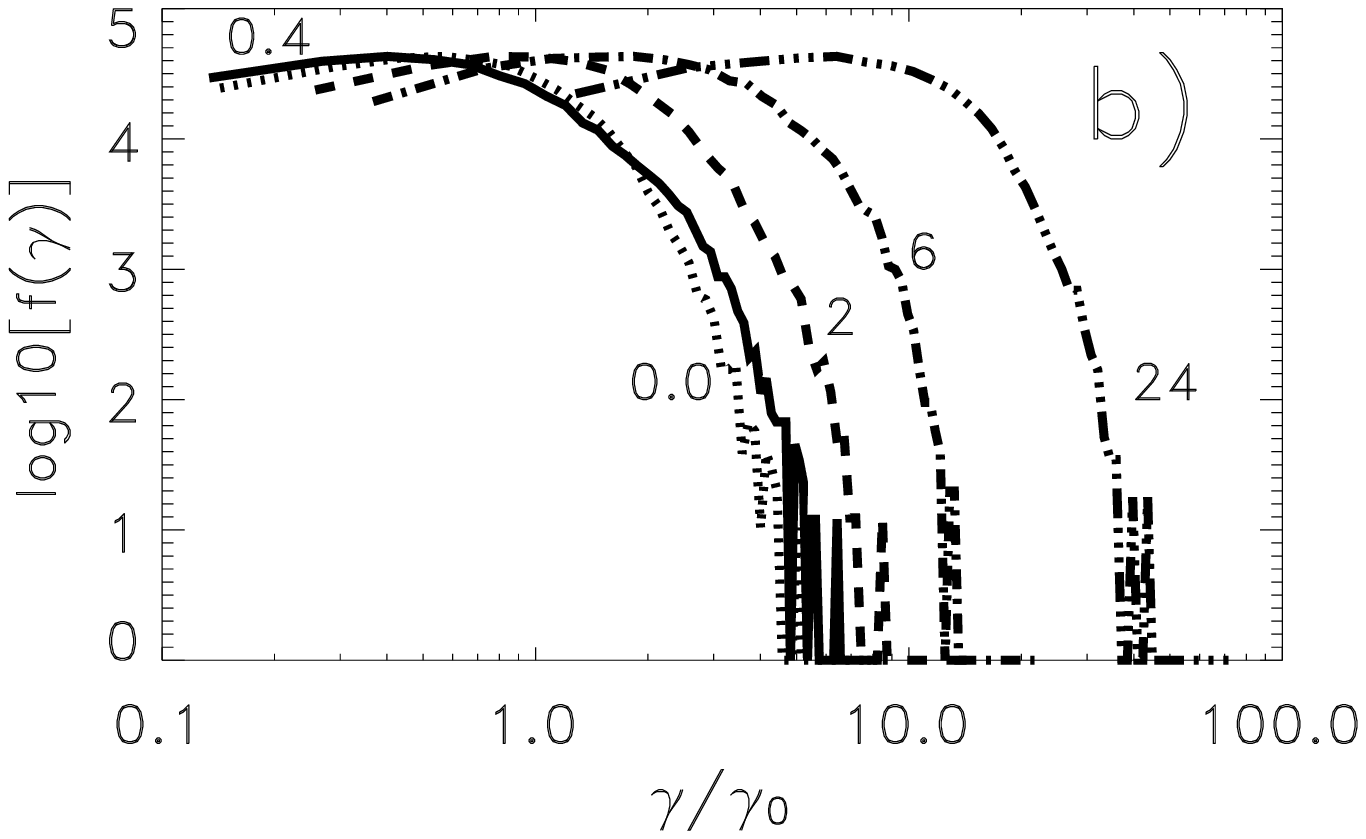}
\caption{The energy distribution $f(\gamma)$ of positrons
for $8 < \xdimless < 16$ ({\it a}) and $50 < \xdimless < 58$ ({\it b}), 
normalized such that $\int{d\gam} f(\gam) = n^+$. 
The labels denote $\sigmaant$; the 
other parameters are as in Fig. \ref{fig:one}.}
\lab{fig:gammadist}
\lab{fig:two}
\efig

\bfig
\incg[width=0.5\textwidth]{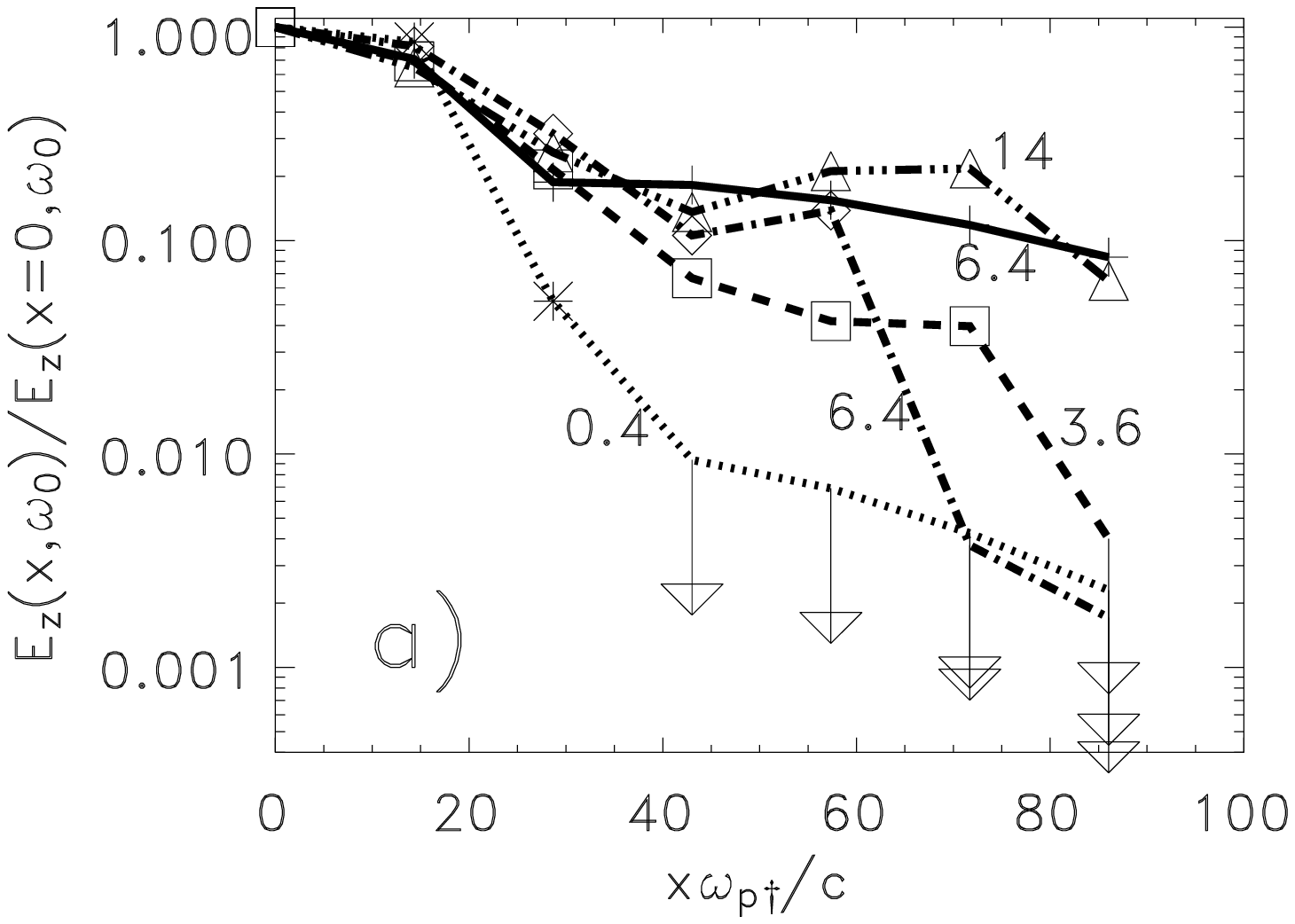}
\incg[width=0.5\textwidth]{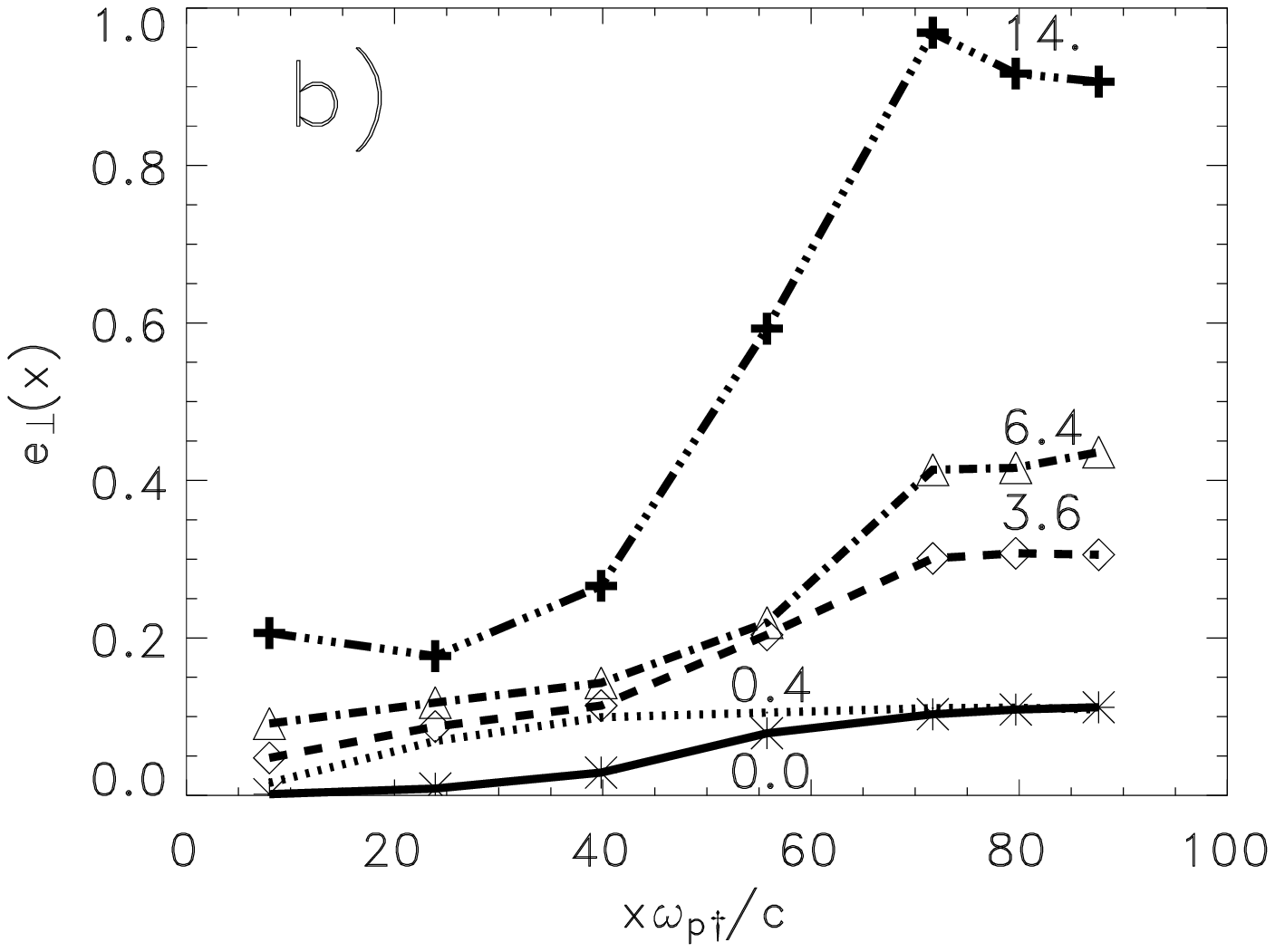}
\caption[]{{\it (a)}
Fourier amplitude of the TEM wave at $\omega=\omegaant$ as a function
of $\xdimless$, normalized to its value at $\xdimless = 0$.
The solid line represents free propagation (no shock). 
The other curves are for shocked TEM waves.
Arrows indicate upper limits, and labels denote $\sigmaant$.
{\it (b)} Transverse momentum density 
$\eperp(\xdimless) = n(\xdimless)\uperp(\xdimless)/c$, 
in arbitrary units.
As the wave decays, $\eperp$ increases due to plasma heating.
Each curve represents a shocked TEM wave.
Both panels are snapshots at $t\simeq \tref + 100\omegapd^{-1}$.
Other parameters are as in Fig. \ref{fig:one}.}
\lab{fig:ldecay}
\lab{fig:three}
\efig

\bfig
\incg[width=0.5\textwidth]{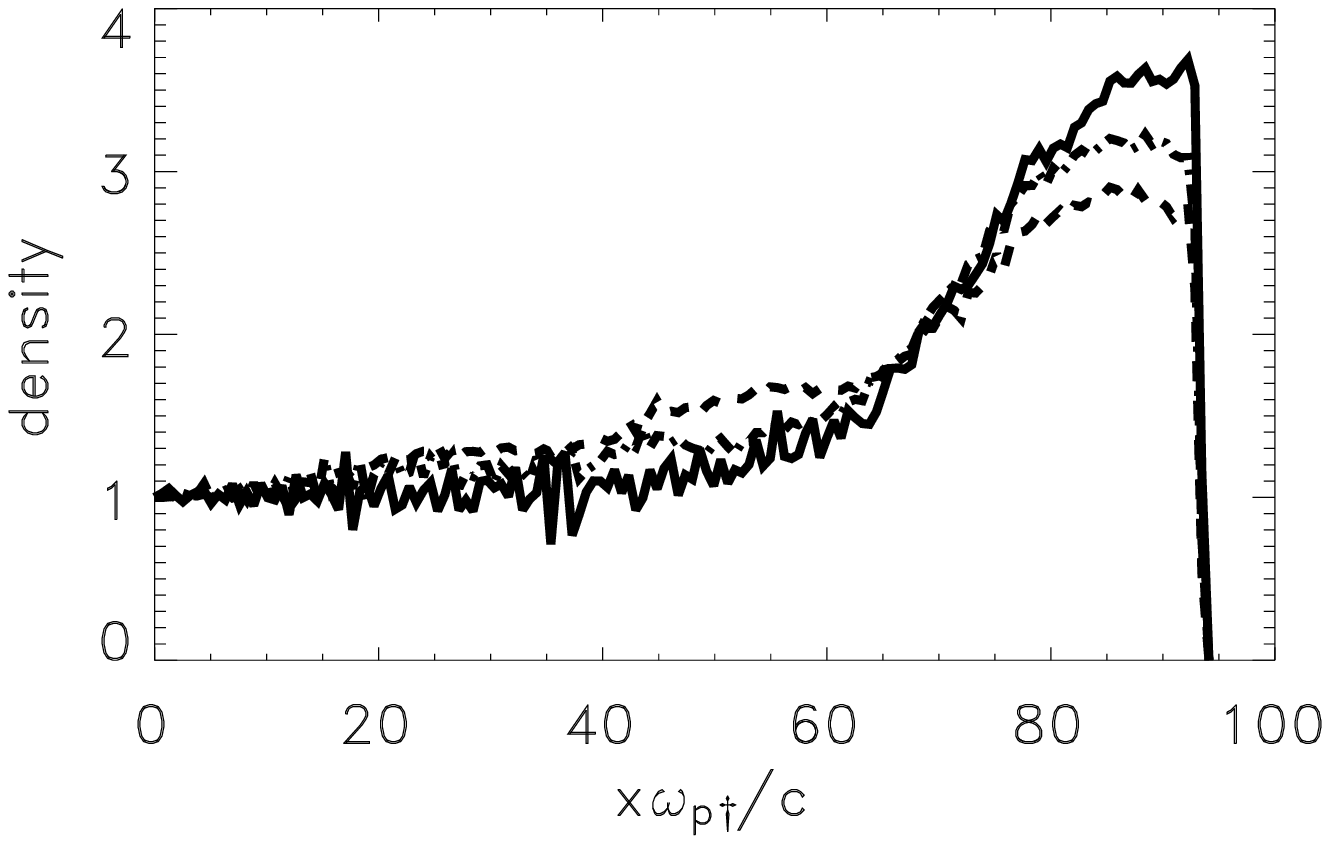}
\caption[]{The aggregate density profile 
$n(\xdimless)=n^+(\xdimless) + n^-(\xdimless)$ for a TEM wave shock. 
The regions $10 \lesssim \xdimless \lesssim 50$, 
$50 \lesssim \xdimless \lesssim 80$,
and $\xdimless \gtrsim 80$ are the shock precursor, 
the shock interior and the downstream medium, respectively (see $\S 2$).
The drop at $\xdimless \gtrsim 95$ is due to the magnetic wall.
The solid, dashed-dotted, and dashed lines are for 
$\sigmaant= 24, \sigmaant=2.0$, and $\sigmaant=0.4$, respectively. 
Other parameters are as in Fig. \ref{fig:one}.}
\lab{fig:profiles}
\lab{fig:four}
\efig

\end{document}